\documentclass[english]{article}
\usepackage[T1]{fontenc}
\usepackage[latin9]{inputenc}

\makeatletter
\newcommand{\lyxaddress}[1]{
\par {\raggedright #1
\vspace{1.4em}
\noindent\par}
}

\makeatother

\usepackage{babel}
\begin{document}

\title{\textbf{The YARK theory of gravity can reproduce neither the LIGO\textquoteright s
\textquotedblleft GW150914 signal\textquotedblright{} nor the other
LIGO's detections of gravitational waves}}

\author{\textbf{Christian Corda}}
\maketitle

\lyxaddress{\textbf{International Institute for Applicable Mathematics and Information
Sciences, B. M. Birla Science Centre, Adarshnagar, Hyderabad 500063
(India); e-mail address: cordac.galilei@gmail.com}}
\begin{abstract}
We show that, based on important reasons, differently from the some
recent claim in the literature, the YARK theory of gravity can reproduce
neither the LIGO's \textquotedblleft GW150914 signal\textquotedblright{}
nor the other LIGO's detections of gravitational waves (GWs).
\end{abstract}
The self-called (from the initials of the proper surnames of its authors)
YARK theory of gravity was originally proposed by T. Yarman in Foundation
of Physics \cite{key-1}. After that, various papers on the YARK theory
of gravitation have been published by T. Yarman and collaborators
(O. Yarman, A. L. Kholmetskii and M. Arik. Hereafter we will refer
to them as the YARK group) {[}3 - 10{]}. This is a bit surprising,
because the YARK theory of gravity is a non-metric theory and, as
we have rigorously shown in our recent paper \cite{key-11} (but this
issue is well known, see for example Will's review \cite{key-12}),
it macroscopically violates Einstein's equivalence principle, which
has today a strong, indisputable, empiric evidence \cite{key-11,key-12}.
On the other hand, despite the YARK group claims that their theory
is in agreement with various experimental and/or observational results
{[}3 - 10{]}, we have shown that it cannot explain the Mössbauer rotor
experiment \cite{key-13}, contrary to their claims in \cite{key-5}.
In fact, in \cite{key-13,key-14} we have shown that in the Mössbauer
rotor experiment an additional effect of clock synchronization is
present, which has not been considered by the YARK group in \cite{key-5}.
Remarkably, our result on the Mössbauer rotor experiment can be considered
a new proof of Einstein's general theory of relativity, and, for this
reason, it has been recently awarded with an Honorable Mention in
the 2018 Essay Competition of the Gravity Research Foundation \cite{key-15}. 

For the sake of completeness, it is better adding some detail on the
YARK theory of gravity. It is basically based on the energy conservation
law \cite{key-5}. In such an approach, space-time is Minkowskian-like
and the Christoffel symbols are equal to zero. In addition, the diagonal
metric coefficients are multiplied by a conformal factor that depends
on the ``static gravitational binding energy'' of a test particle
\cite{key-5}. Thus, the YARK group claims that their theory combines
both metric and dynamical approaches \cite{key-5}. This seems an
apparent contradiction with our above statement that \textquotedbl{}the
YARK theory of gravity is a non-metric theory\textquotedbl{}. But
the key point is that, instead of the geodesic motion in a curved
space-time, the YARK theory shows a dynamical equation of motion for
a test particle in a flat space-time governed by the force resulting
from the spatial variation of the gravitational binding energy \cite{key-5}.
But in the common language of gravitational theorists, a metric theory
is defined as being a theory which satisfies Einstein's equivalence
principle \cite{key-12}, which, in turn, implies that gravitation
must be a \textquotedblleft curved space-time\textquotedblright{}
phenomenon \cite{key-12}. This means that the effects of gravitation
are completely equivalent to the effects of living in a curved space-time
\cite{key-12}. Thus, we have at least two issues in the YARK theory
of gravity which are in strong contrast with Einstein's equivalence
principle. The first is that the space\textendash time is flat \cite{key-5}.
The latter which states that the gravitational energy can be localized
\cite{key-5}. For this second point, we recall that another consequence
of Einstein's equivalence principle is indeed that one can always
find in any given locality a reference's frame (the local Lorentz
reference's frame) in which ALL local gravitational fields are null.
No local gravitational fields means no local gravitational energy-momentum
and, in turn, no stress-energy tensor for the gravitational field
\cite{key-11,key-19}.

Another recent claim of the YARK group is that their theory can reproduce
the LIGO's \textquotedblleft GW150914 signal\textquotedblright{} as
well as the other recent LIGO's GW detections \cite{key-10}. Here
we show that this claim is not correct. In fact, we see that in \cite{key-10}
(in the paper \cite{key-10} C.B. Marchal was an additional author)
the YARK group claims that, while GWs are not present in their theory,
the same theory admits a difference $\triangle\varphi$ of the phase
shift of light due to electromagnetic radiation incoming from the
coalescence of super-massive bodies in a distant binary system, see
equations (11) and (14) in \cite{key-10}. In the YARK theory, such
a value $\triangle\varphi$ is the same for both of the arms of the
interferometer and the ratio $|\frac{\triangle\varphi}{\varphi}|$,
where $\varphi$ is the phase of the interferometer light beam, should
be of the same order of magnitude of the GW150914 signal \cite{key-10}.
But the key point is that LIGO and the other interferometric GW detectors
operate in a \emph{differential mode}. In other words, the total phase
shift of light, which represents the output of LIGO, is given by \emph{the
difference} between the phase shifts of light of each arm of the interferometer,
see for example \cite{key-16,key-17}. Thus, if the value $\triangle\varphi$
in the YARK theory is the same for both of the arms of the interferometer,
the total output of LIGO must be null. In other words, if the YARK
theory is correct, LIGO cannot detect signals.

On the other hand, in \cite{key-10} the YARK group claims that the
relative phase shift which is admitted by their theory has electromagnetic
origin. But this results in contrast with the event GW170817 \cite{key-18}.
In fact, in such an event there was a time delay in the detection
of the electromagnetic counterpart of the LIGO signal \cite{key-18}.
 Thus, if the YARK theory is correct, it should have been impossible
having such a delay.

Resuming, in this work it has been shown that, based on important
reasons, differently from the claims in \cite{key-10}, the YARK theory
of gravity can reproduce neither the LIGO's \textquotedblleft GW150914
signal\textquotedblright{} nor the other LIGO's detections of GWs.
If we add that the YARK theory of gravity violates Einstein's equivalence
principle \cite{key-11} and is in disagreement with the Mössbauer
rotor experiment {[}14 - 16{]}, we conclude that this theory is largely
disfavoured. 

\section*{Acknowledgements}

The author thanks the referees of this paper for useful comments.

\end{document}